\documentclass{aa}
\usepackage{epsfig}
\usepackage{psfig}
\usepackage{graphicx}
\usepackage{txfonts}

\usepackage{lscape}

\titlerunning{Proper Motion of RX J0822$-$4300}
\newcommand{\rx}{RX J0822$-$4300}
\newcommand{\pu}{Puppis$-$A}
\newcommand{\RX}{RX~J0822$-$4300}
\newcommand{\Pu}{Puppis$-$A}

\begin{document}
\title{Probing the Proper Motion of the Central Compact Object in 
       Puppis-A with the Chandra High Resolution Camera} 

\author{C. Y. Hui \and W. Becker}   
\date{Received 30 June 2006 / Accepted 8 August 2006} 
\institute{Max-Planck Institut f\"ur Extraterrestrische Physik, 
          Giessenbachstrasse 1, 85741 Garching bei M\"unchen, Germany}

\abstract{
 By making use of the sub-arcsecond angular resolution of the High Resolution Camera (HRC-I)
 aboard the Chandra X-ray satellite we have examined the central compact object \rx\, in the 
 supernova remnant Puppis-A for a possible proper motion. Using data which span an epoch 
 of 1952 days we found the position of \rx\, different by $0.574\pm0.184$ arcsec, implying 
 a proper motion of $\mu=107.49\pm34.46$ mas/yr. For a distance of 2.2 kpc, this proper 
 motion is equivalent to a recoil velocity of $1121.79\pm 359.60$ km/s. The position angle 
 is found to be $241^\circ \pm 24^\circ$. Since both the magnitude and direction of the proper motion
 are in agreement with the birth place of \rx\, being near to the optical expansion 
 center of the supernova remnant, the result presented in this letter is a promising indication 
 of a fast moving compact object in a supernova remnant. Although the positional shift inferred 
 from the current data is significant at a $\sim3\sigma$ level only, one or more future HRC-I observations 
 can obtain a much larger positional separation and further constrain the measurement. 

 }

\maketitle

\vspace{-1.5cm}
\section{Introduction}
 The group of supernova remnant (SNRs) which are known to host a radio-quiet but X-ray 
 bright central compact object (CCO) is a slowly growing one. Thanks to more 
 sensitive X-ray observatories it currently 
 includes Cas$-$A (CXOU J232327.9+584842; Tananbaum 1999), RX J0852.0$-$4622 
 (CXOU J085201.4$-$461753; Aschenbach 1998), RX J1713.7$-$3946 (RX J1713.4$-$3949; 
 Pfeffermann \& Aschenbach 1996), RCW\,103 (1E 161348-5055; Tuohy \& Garmire 1980), 
 \Pu\,(\rx; Petre et al.~1982; Petre, Becker, \& Winkler ~1996; Hui \& Becker 2006), 
 PKS 1209-51/52  (1E 1207.4$-$5209; Helfand \& Becker 1984) and Kes 79 
 (CXOU J185238.6$+$004020; Seward et al.~2003; Gotthelf et al.~2005).

 \rx\, was first noticed in one of the EINSTEIN HRI images of the \pu\ supernova remnant 
 (Petre et al.~1982). However, it appeared in this data only as a faint X-ray feature. It 
 is not until the era of ROSAT when it became evident that \rx\, is the central compact 
 stellar remnant which was formed in the supernova event (Petre, Becker, \& Winkler 1996). 

 Recently, Hui \& Becker (2006) presented results from a detailed analysis of \rx\, 
 which made use of all XMM-Newton and Chandra data available from it by beginning of 2005. 
 The spectral analysis of XMM-Newton data revealed that the X-ray emission from \rx\, 
 is in agreement with being of thermal origin. A double blackbody model with the temperatures 
 $T_{1}=(2.35-2.91)\times10^{6}$ K, $T_{2}=(4.84-5.3)\times10^{6}$ K and the projected 
 blackbody emitting radii $R_{1}=(2.55-4.41)$ km and $R_{2}=(600-870)$ m gave the best 
 description of the observed point source spectrum among various spectral models tested. 
 The X-ray images taken with the Chandra HRC-I camera allowed for the first time to 
 examine the spatial nature of \RX\, with sub-arcsecond resolution. Besides an accurate  
 measurement of the source position, this observation constrained the point 
 source nature of \rx\, down to $0.59\pm0.01$ arcsec (FWHM) for the first time.

 Despite the effort in searching for coherent radio pulsations, \rx\, has not been detected
 as a radio pulsar (Kaspi et al. 1996). Similar to many other CCOs, it has no optical
 counterpart down to a limiting magnitude of $B\ga 25.0$ and $R\ga 23.6$ (Petre, Becker,
 \& Winkler 1996). Together with the lack of long term flux variation (Hui \& Becker 2006),
 all these evidences rule out many types of X-ray sources as a likely counterpart of
 \rx, except a neutron star.

 \rx\, is located about $\sim 6$ arcmin distant from the optical expansion center of \pu, 
 which is at RA=$08^{\rm h}22^{\rm m}27.45^{\rm s}$ and Dec=$-42^{\circ}57'28.6"$ (J2000)
 (cf.~Winkler \& Kirshner 1985; Winkler et al.~1988). The age of the supernova remnant, 
 as estimated from the kinematics of oxygen-rich filaments is $\sim 3700$ years (Winkler 
 et al.~1988). If these estimates are correct, \rx\, should have a proper motion of 
 the order of $\sim 100$ mas/yr to a direction away from its proposed birth place. 

 In this letter we test this hypothesis by making use of archival Chandra HRC-I data 
 spanning an epoch of somewhat more than five years. The expected positional displacement 
 for \RX\ in this time span is of the order of $\sim 0.5$ arcsec. This is in the range of 
 the Chandra accuracy given the possibility to correct for pointing uncertainties by using
 X-ray counterparts of stars with accurate position and which are serendipitously located 
 in the field of view. 

\vspace{-0.5cm}
\section{Observation and data analysis} 

 Owing to the fact that there are only few bad pixels in the Chandra HRC-I and the pixel 
 size of 0.13187 arcsec oversamples the point spread function (PSF) by a factor of $\sim 5$, 
 the HRC-I appears to be the most suitable detector to perform astrometric measurements of 
 X-ray sources. 

 Checking the Chandra archive for suitable data we found that by mid of 2006 \rx\, was 
 observed twice with the HRC-I. The first observation was performed in 1999 December 21 
 (MJD 51533) for an exposure time of about 16 ksec. The second observation was done in 
 2005 April 25 (MJD 53485) for an on-source time of $\sim 33$ ksec. In the April 2005 
 observation, the target was displaced only $\sim 0.2$ arcmin off from the optical axis of 
 the X-ray telescope. In the December 1999 data, it appears to be $\sim 0.3$ arcmin 
 off-axis which, in both cases, is small enough to have a negligible effect for the 
 distortion of the PSF relative to an on-axis observation.
 
 In order to increase the precision required for accurate astrometric measurements, 
 systematic uncertainties need to be corrected. Apart from the aspect offset 
 correction we also considered the errors introduced by determining the event positions.
 The later included corrections of the tap-ringing distortion in the position reconstruction 
 and the correction of errors introduced by determining the centroid of the charge cloud.
 Instead of using the fully processed pipeline products, we started our data reduction 
 with level-1 files to be able to correct for these systematic effects. All the data 
 processings were performed with CIAO 3.2.1. Details of the applied corrections are 
 described in the following. 
 
 Instabilities in the HRC electronics can lead to a tap-ringing distortion in the
 position reconstruction of events. Correction has been applied to minimize this 
 distortions in standard HRC level-1 processing which required to know the values 
 of the amplitude scale factor (AMP\_SF). Such values are found in the HRC telemetry and 
 are different for each event. Unfortunately, they are often telemetered incorrectly. 
 In order to fix this anomaly, we followed a thread in CIAO to deduce the correct 
 values of AMP\_SF in the level-1 event file from other HRC event data and applied 
 these corrected values to minimize the distortion. 

 The de-gap correction was applied to the event files in order to compensate
 the systematic errors introduced in the event positions by the algorithm which
 determines the centroid of the charge cloud exiting the HRC rear micro-channel 
 plate. 

 After correcting these systematic errors we generated the level-2 event lists 
 files which were used thoroughly for the remaining analysis. We created HRC-I 
 images of \rx\, for both epochs with a binning factor of 1 so that each pixel 
 has a side length of 0.13187 arcsec.

 To be able to correct for pointing uncertainties by using X-ray counterparts of 
 stars which have their position known with high accuracy, we applied a wavelet 
 source detection algorithm to the HRC-I images. Two X-ray point sources with 
 a count rate of about 1\% and 3\% relative to that of \rx\, were detected 
 serendipitously at about $\sim2.5$ arcmin and $\sim5.5$ arcmin distance from \rx. Figure 1 
 shows a $9.5\times 7$ arcmin field surrounding \rx\, as seen with the HRC-I in April 2005. 
 Both serendipitous sources, denoted as A and B, are indicated in this figure. 
 Their X-ray properties are summarized in Table 1.  

 \begin{table*}
 \caption{X-ray properties of serendipitous sources in the neighborhood of \rx.}
 \centering
 \begin{tabular}{lcccc}
 \hline\hline
Source & \multicolumn{2}{c}{X-ray position} & Positional error$^a$ & Net count rate \\
       & RA (J2000) & Dec (J2000) &   & cts/s \\
  \hline\\[-1.5ex]
 \multicolumn{5}{c}{{\bf 1999 observation}}\\[1ex]
 \hline
  A & $08^{\rm h}21^{\rm m}46.339^{\rm s}$ & $-43^{\circ}02'03.73"$ & 0.16" & $(2.51\pm 0.38)\times 10^{-3}$ \\
  B & $08^{\rm h}22^{\rm m}23.924^{\rm s}$ & $-42^{\circ}57'59.29"$ & 0.20" & $(8.42\pm 0.75)\times 10^{-3}$ \\
  \hline\\[-1.5ex]
  \multicolumn{5}{c}{{\bf 2005 observation}}\\[1ex]
  \hline
  A & $08^{\rm h}21^{\rm m}46.313^{\rm s}$ & $-43^{\circ}02'03.46"$ & 0.07" & $(3.20\pm 0.30)\times 10^{-3}$\\
  B & $08^{\rm h}22^{\rm m}23.966^{\rm s}$ & $-42^{\circ}57'59.70"$ & 0.18" & $(7.47\pm 0.48)\times 10^{-3}$ \\
    \hline
  \end{tabular}
  \\
$^a$ The positional errors of sources A and B are given by the PSF fitting and the wavelet detection algorithm 
respectively.
\end{table*}

 In order to determine their X-ray positions with higher accuracy than possible
 with the wavelet analysis, we fitted a 2-D Gaussian model with the modeled PSF 
 as a convolution kernel to both sources A and B. These fits require some 
 information on the source energy spectrum which is not available from the HRC-I 
 data. We therefore checked the archival XMM-Newton data for both sources and 
 found from an spectro-imaging analysis of MOS1/2 data (cf.~Figure 1 in Hui \& Becker 
 2006) that the hardness-ratios of source A and B are comparable to that of \rx\, 
 which has its energy peak at $\sim 1.5$ keV. We therefore extract the 
 Chandra HRC-I PSF model images at 1.5 keV with the corresponding off-axis angles 
 from the CALDB standard library files 
 (F1) by interpolating within the energy and off-axis grids using the CIAO tool 
 MKPSF. The exposure maps were also generated at this energy by using MKEXPMAP. 
 
 The size of the $1-\sigma$ error circles of source A obtained by this method are 
 0.16 arcsec and 0.07 arcsec for the 1999 and 2005 observations, respectively. 
 The relatively large error in the December 1999 observation is due to its shorter 
 integration time and thus smaller photon statistics.

 The large off-axis angle ($\sim 5.5$ arcmin) of source B causes a marked blurring 
 of the PSF (90\% encircled energy radius $\simeq 4$ arcsec). Such distortion makes 
 the source appear to be very dispersed. Given the patchy and uneven supernova 
 remnant emission this source is surrounded by and the limited photon statistics  
 we did not succeed in obtaining its coordinates more accurate than possible with the
 wavelet algorithm (which is $\sim 0.2$ arcsec). This leaves source A as the only
 reference star to perform astrometric correction.

 Correlating the source position of source A with the Two Micron All Sky Survey (2MASS) catalog 
 (Skrutskie et al.~2006) identified the star with the source designation 08214628$-$4302037  
 as a possible optical counterpart. Since the next nearest optical source is located 
 about 5 arcsec away from the X-ray position of source A, we adopt the 2MASS source 
 08214628$-$4302037 as its optical counterpart. The visual magnitudes of the object 
 are $J=12.161\pm 0.027$, $H=11.675\pm 0.023$ and $K=11.558\pm 0.024$. 
 Since its spectral type is not known with certainty, we adopted a typical X-ray-to-optical 
 flux ratio of log$(F_x/F_{opt})\simeq -2.46$ for stars from  Krautter et al.~(1999). 
 Assuming a Raymond-Smith thermal plasma model with $kT=0.15$ keV, $n_H=4\times10^{21}$ 
 cm$^{-2}$ (Hui \& Becker 2006) and solar abundances for the star's spectrum we estimated
 with the aid of PIMMS (version 3.8a2) its HRC-I count rate to be $\sim 3\times10^{-3}$ 
 cts/s. This is in good agreement with the observed count rate of source A (cf.~Table 1).

 In order to use the optical identification of the serendipitous X-ray source A as a reference source 
 for the offset correction, we have to check whether itself shows a proper 
 motion. To investigate this, we correlated its 2MASS position with the UCAC2 catalog 
 (Zacharias et al. 2003). We unambiguously found a source with the UCAC2 designation 13302738 
 as a counterpart of the X-ray source A. According to this catalog, this source has a proper motion 
 of $\mu_{\rm RA}=-16.0\pm5.2$ mas/yr, $\mu_{\rm dec}=-1.7\pm5.2$ mas/yr. 

 We attempted to make an independent estimate by 
 analysing the images from the first and the second 
 Digitised Sky Surveys\footnote{http://ledas-www.star.le.ac.uk/DSSimage/}. 
 From the observation dates specified for the DSS-1 and DSS-2 images, 
 we found that the epoches of these two images are separated by 5134 days.
 We took four bright stars within 1 arcmin neighbourhood of the X-ray source A
 as the references to align the frames of DSS-1 and DSS-2. 
 None of these stars appeared to be saturated so that their 
 positions could be properly determined by a 2-D Gaussian fit. 
 We determined the offset between these two images from comparing the
 best-fit positions of the four reference stars in both frames. However, we 
 found that the alignment error is at a level of $\sim0.5$ arcsec. This is 
 close to the average positional discrepancy between DSS-1 and DSS-2 which 
 is found to be $\sim0.6$ arcsec. The information provided by DSS-1/2 
 thus does not allow us to estimate the proper motion of our source of interest 
 independently.  In view of this, we can only resort on the findings in the 
 UCAC2 catalog. 

 Under the assumption that the UCAC2 object 13302738  
 is indeed the optical counterpart of source A, we applied the aspect 
 correction to the corresponding frames with the proper motion of the reference star 
 taken into account. However, with only one comparison 
 source avaliable for the frame alignment, there are some limitations in our 
 adopted method. Firstly, the roll angle between two frames cannot be 
 determined independently with just one reference source. Hence, the accuracy 
 of the current result is limited by the output of the star-tracking camera, the 
 Pointing Control and Determination system (PCAD). Also, an independent 
 estimate of the HRC-I plate scale cannot be made with only one reference 
 source. The potential variation of the plate scale might introduce an extra 
 error, though we consider this is negligible as the typical uncertainty of 
 the HRC-I plate scale is at the order of $\sim0.05$ mas/pixel
 \footnote{http://cxc.harvard.edu/cal/Hrma/optaxis/platescale/geom\_public.html}.

 The error circle of the UCAC2 object 13302738 is specified to be 0.015 arcsec. Including the uncertainty in the 
 proper motion, the overall positional error of this astrometric source is increased with time which gives 
 0.016 arcsec and 0.037 arcsec in the December 1999 and April 2005 epoch respectively. 
 The total error of the aspect corrections is calculated by combining the statistical error of the X-ray position of 
 source A and the astrometric error of the UCAC2 object in quadrature. This yields a $1-\sigma$ 
 error of 0.161 arcsec and 0.079 arcsec for the aspect correction of the December 1999 and April 2005 
 observation, respectively. 

 The position of \rx\, was determined by the same procedure we applied to obtain
 the position for source A. The fits provide 
 us with the coordinates for \rx\, which are RA=$08^{\rm h}21^{\rm m}57.389^{\rm s}$ 
 and Dec=$-43^{\circ}00'16.90"$ (J2000) in the 1999 observation and 
 RA=$08^{\rm h}21^{\rm m}57.343^{\rm s}$ and Dec=$-43^{\circ}00'17.18"$ (J2000)
 in the April 2005 observation\footnote{The best-fitting position in the 1999 observation 
 has a 0.242 arcsec deviation from that inferred by Hui \& Becker (2006) which 
 corrected the offset by simply using the calculator provided by the Chandra Aspect team  
 http://cxc.harvard.edu/cal/ASPECT/fix\_offset/fix\_offset.cgi}.
 The  $1-\sigma$ error of the position introduced by the PSF-fit is 
 0.01 arcsec in both epochs. 
 In order to exclude any dependence of the deduced source positions on the aperture size
 of the selected source region, we repeated the fits for three different apertures with
 radii equal to 3 arcsec, 4 arcsec and 5 arcsec, respectively. We did not find any
 variation of the best-fitted parameters in these independent fittings.

\begin{figure}
\psfig{figure=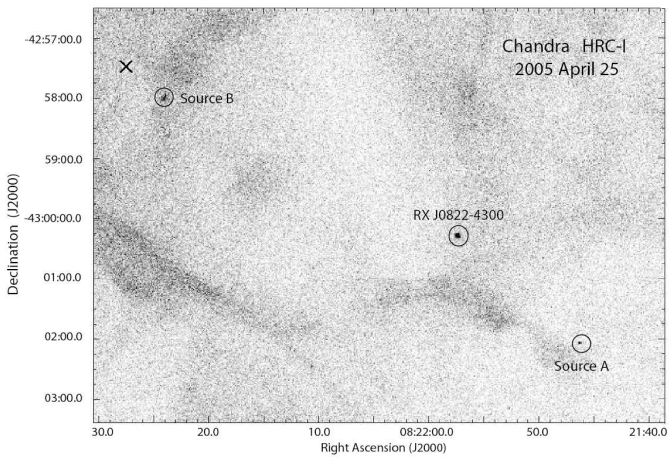,width=9cm,clip=}
\caption{Part of the Chandra HRC-I image of the \pu\ supernova remnant. Besides \rx, there are 
 two serendipitous point sources detected in the field of view. They are denoted as source A and 
 B. The optical expansion center of \pu\ as obtained by Winkler et al.~(1988) is marked by a cross. 
}
\end{figure}

\vspace{-0.5cm}
\section{Results and Discussion}
 The fitted X-ray positions of \rx\, are plotted for the observational epochs 
 December 1999 and April 2005 in Figure 2. The size of the error circle for each position 
 is determined by adding the error in correcting the aspect offset and the error of the 
 position of \rx\, in quadrature. This gives 0.162 arcsec and 0.088 arcsec for the December 
 1999 and the April 2005 observations, respectively. The corresponding error circles are
 indicated in Figure 2.  

 Comparing the positions of \rx\, as deduced for the two observations in 1999 and 2005
 we found that they are different by $0.574\pm 0.184$ arcsec. This is well consistent with the displacement
 estimated from the kinematic age of the SNR and the positional offset of \rx\, from the
 SNR's optical expansion center (cf.~\S1). The quoted error is $1-\sigma$ and combines 
 the positional errors in both epochs in quadrature. 
 
 Given the epoch separation of 5.34 years for both Chandra HRC-I observations the observed 
 proper motion of \rx\, is $\mu=107.49\pm 34.46$ mas/yr. The position angle is PA=$241^{\circ}\pm24^{\circ}$.  
 Taking the position inferred from the December 1999 observation and the position of the
 optical expansion center of \pu\, (Winkler et al.~1988), a position angle of 
 PA$\simeq 243^{\circ}$ is implied. Such consistency suggests that \rx\, is indeed physically 
 associated with \pu\, and its actual birth place is not too far away from the SNR's optical 
 expansion center. Assuming that the distance of 2.2 kpc to \pu\, is correct, the observed proper motion 
 implies that \rx\, has a projected recoil velocity as high as $1121.79\pm 359.60$ km/s to the 
 southwest. The inferred transverse velocity seems high compared with the average velocity, $\sim 250$ km/s,
 of ordinary radio pulsars (Hobbs et al. 2005) though the large error prevents any constraining conclusions.
 We would like to mention, though, that there are several fast moving pulsars including PSRs B2011+38 and 
 B2224+64 which have 2-D speeds of $\sim 1600$ km/s (Hobbs et al. 2005).

 The interesting association of \pu\ and \rx\, provides a unique opportunity to observe the 
 interaction between a compact stellar remnant and a SNR. Also, this system is a test-bed 
 for studying the dynamics of the supernova explosion. Measuring the proper motion of \rx\,
 with higher accuracy than possible with the current data will make it possible 
 to provide constrains for the SN explosion model which formed \pu. 

 Although our result is a promising indication of a fast moving CCO in a SNR,  we note that the 
 deduced positional shift is only significant at a $3.1\sigma$ level. 
 Since the X-ray source A is rather faint, its relatively large positional error predominates in the error budget.

 \begin{figure}
 \centerline{\psfig{figure=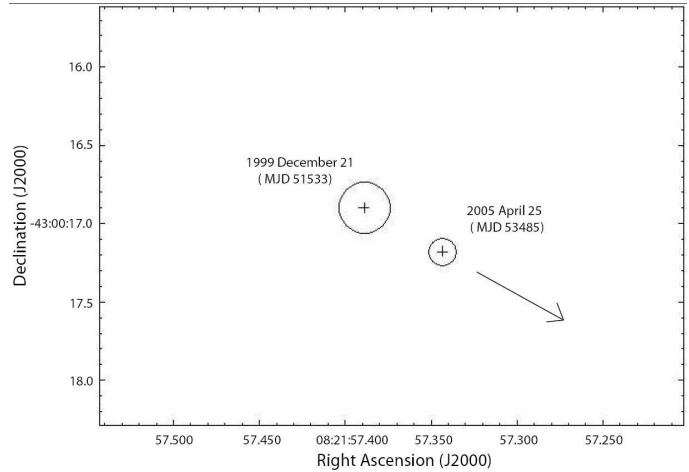,width=9cm,clip=}}
 \caption{The best-fitted X-ray positions of \rx\, in two epochs are marked by crosses.
  The circles indicate the $1-\sigma$ error. 
  The arrow shows the direction of proper motion inferred from both positions.
     }
 \end{figure}

 In this letter we have presented the first attempt to measure the proper motion 
 of \rx. In order to further constrain this measurement, one or more Chandra 
 HRC-I observations can attain a much larger positional separation in a few years from today. 
 Furthermore, a more accurate proper motion measurement for the optical counterpart of the reference source 
 is also necessary.

\begin{acknowledgements}
We would like to thank the referee for thoroughly reading the manuscript and provide us with many useful comments.
\end{acknowledgements}

\end{document}